\documentclass[aps,nofootinbib,showpacs,twocolumn]{revtex4}

\usepackage{amsmath, amsfonts, amssymb, dsfont}
\usepackage{graphicx, epsfig, epstopdf}
\usepackage{hyperref}                            

\newcommand{\idMatrix}{\mathds{1}}
\newcommand{\pd}{\partial}
\newcommand{\ii}{\mathrm{i}}

\newcommand{\diff}{\mathrm{d}}

\newcommand{\Angstrom}{\text{\AA}}
\newcommand{\Reals}{\mathbb{R}}
\DeclareMathOperator{\sgn}{sgn}          
\DeclareMathOperator{\tr}{tr}            
\DeclareMathOperator{\RRe}{\mathrm{Re}}  
\DeclareMathOperator{\IIm}{\mathrm{Im}}  
\newcommand{\hc}{^\dagger}               
\newcommand{\cc}{^\ast}               
\newcommand{\avg}[1]{\langle{#1}\rangle} 
\newcommand{\bvec}[1]{\mathbf{#1}}                     

\renewcommand{\thesection}{\arabic{section}}
\newcommand{\jetlsection}[1]{\stepcounter{section}\textbf{\thesection.~{#1}}}

\newcommand{\LL}{\text{L}}
\newcommand{\RR}{\text{R}}
\newcommand{\FF}{\text{F}}

\begin{document}

\title{Effects of quantum recoil forces in resistive switching in memristors}
\author{Oleg~G.~Kharlanov} 
\affiliation{Faculty of Physics, Lomonosov Moscow State University, 1/2 Leninskie Gory, 119991 Moscow, Russia}

\pacs{05.60.Gg, 73.63.-b, 73.63.Rt}

\begin{abstract}
    Memristive devices, whose resistance can be controlled by applying a voltage and further retained, are attractive as possible circuit
    elements for neuromorphic computing. This new type of devices poses a number of both technological and theoretical challenges. Even the physics
    of the key process of resistive switching, usually associated with formation or breakage of conductive filaments in the memristor, is not
    completely understood yet. This work proposes a new resistive switching mechanism, which should be important in the thin-filament regime and
    take place due to the back reaction, or recoil, of quantum charge carriers---independently of the conventional electrostatically-driven ion
    migration. Since thinnest conductive filaments are in question, which are only several atoms thick and allow for a quasi-ballistic, quantized
    conductance, we use a mean-field theory and the framework of nonequilibrium Green's functions to discuss the electron recoil effect for a
    quantum current through a nanofilament on its geometry and compare it with the transmission probability of charge carriers. Namely, we first
    study an analytically tractable toy model of a 1D atomic chain, to qualitatively demonstrate the importance of the charge-carrier recoil, and
    further proceed with a realistic molecular-dynamics simulation of the recoil-driven ion migration along a copper filament and the resulting
    resistive switching.  The results obtained are expected to add to the understanding of resistive switching mechanisms at the nanoscale and to
    help downscale high-retention memristive devices.
\end{abstract}

\maketitle

\jetlsection{Introduction.} Memristors---electric devices whose resistance ``remembers'' the history of the previously passed
current~\cite{Chua_Memristor}---have attracted a lot of theoretical and experimental attention over the past decade, notably since the first
working memristive device was manufactured in 2008~\cite{Memristor2008}. Memristors are particularly appealing in the context of their possible
use for in-memory computing and hardware neural networks, as opposed to the von Neumann architecture~\cite{Ielmini_inmemory}. Current-controlled
resistive switching in such devices typically occurs due to ion migration in a dielectric layer between the anode and the cathode resulting in
the growth or rupture of conductive filaments~\cite{Wang_memristors_review}. Interestingly, despite considerable progress made on the
technological side, there are still gaps in the understanding of the physical processes underlying resistive switching and retention of the
resistive states in memristors. The reason is the stochasticity of the filament dynamics and possible interplay of various scales and effects,
such as Joule heating, surface diffusion, strong electric fields, etc.~\cite{Lee_review, Ielmini_NatComm2019} An improved theory or simulation
frameworks are particularly desirable when one descends to the nanoscale: the constrictions, i.e., the thinnest segments of the conductive
filaments in the corresponding devices, allow for ballistic (quantum) charge transport and exhibit quantized conductance at room temperature,
with the quantization being an attractive property to encode discrete states of the memristor~\cite{Xue_review2019, Gao_Gquant,
Minnekhanov_OrgElec2019, Shvetsov_GquantPPX, Kharlanov2022}. At the same time, interaction of quantum currents with the classical filament and
its molecular environment constitutes a challenging task to describe, limiting rational design of such memristors.

In this letter, by no means aiming at a complete description of the dynamics of quantized-conductance filaments, we analyze a possible role of
electron-induced, ``recoil'' forces acting on the ions \cite{Dundas_NNano2009, Todorov_PRB2010} in these dynamics, which is also potentially
related to the extended lifetimes of filaments with well-quantized conductance values. The point is that conductance quantization in quasi-1D
systems in the units of $G_0 = e_0^2 / \pi\hbar \approx 77.5~\mu\text{S}$ ($e_0$ is the elementary charge, $\hbar$ is the Planck's constant) is
not quite fundamental, being typical of ``smooth'', adiabatic filaments~\cite{Landauer,Buttiker}; in principle, it does not have to hold for
irregular filaments forming in memristors. However, the back reaction of quantum electrons passing through the filament could push it toward
smoother shapes consistent with minimum recoil, integer electron transmission probability (transmittance), and thus quantized conductance.
Indeed, in our recent study~\cite{Kharlanov2022}, we identified that recoil forces should be large enough at voltages $V \sim 1~\text{V}$ and
conductances near $G_0$ to compete with the interatomic forces, potentially affecting the evolution of the filament. Together with that, our
experiments showed that the thinnest filaments with the conductance $G = G_0$ and even $G_0/2$, where the quantum recoil forces should be most
pronounced, demonstrated the longest renention, i.e., stability with time~\cite{Kharlanov2022}. However, Ref.~\cite{Kharlanov2022} described
conduction electrons using a semi-quantitative continuous free-electron model and lacked a dynamical simulation of the filament stochastic
evolution, aiming at an order-of-magnitude estimation of the effects. Moreover, interatomic interactions in the filament were simply reduced to
the latter having a certain amount of surface energy, like a liquid drop. In fact, since ion migration in memristors, e.g., surface diffusion,
proceeds as a sequence of hops~\cite{Ielmini_NatComm2019}, the lack of the dynamics strongly limits the model used in Ref.~\cite{Kharlanov2022},
especially when it comes to simulate an actual resistive switching event. In contrast, in the present work we develop an atomistic quantum
theory of electron recoil forces and incorporate it into a molecular-dynamics (MD) framework with realistic interatomic potentials to observe a
quantum analogue of electromigration~\cite{Electromigration} triggering a resistive switching. The theory and simulations, described in the
following sections, demonstrate that if the conditions for the quantized conductance are not satisfied, resistive switching can be facilitated
by the growing recoil forces. This new resistive switching mechanism, which is important for memristors with atomic-scale filaments and high
current densities, represents the main novelty of the present work, providing an alternative to the conventional mechanism due to ion migration
in the electrostatic field between the cathode and the anode~\cite{Lee_review,Ielmini_NatComm2019,Xue_review2019}. Our results thus indicate
that quantum recoil should be taken into account when describing memristive nanodevices and could potentially be used to achieve better
retention characteristics of their quantized-conductance states.

\jetlsection{The model.} Similarly to earlier studies~\cite{Dundas_NNano2009, Todorov_PRB2010} of current-induced forces, we use a
semiclassical, Born--Oppenheimer (adiabatic) description of the nanoconstriction supporting a quantum current: the positions of the ions are
treated classically, while the conduction electrons are described within the framework of nonequilibrium Green's functions (NEGFs) applied to
the tight-binding model~\cite{Do_NEGFreview}. Namely, the Hamiltonian of the system reads:
\begin{eqnarray}
    \hat{H} &=& \hat{H}^{(\text{el})} + \sum_{i=1}^N\frac{\bvec{P}_i^2}{2M_i} + U(\bvec{X}), \\
    \hat{H}^{(\text{el})} &=& \sum_{\sigma} \int \hat\psi_\sigma\hc(\bvec{x})\Bigl[-\frac{\hbar^2\nabla^2}{2m} +
                                   u(\bvec{x};\bvec{X})\Bigr]\hat\psi_\sigma(\bvec{x}) \diff^3x \nonumber\\
                          &=& \sum_{i,j,\sigma} t_{ij}(\mathbf{X}) \hat{c}_{i\sigma}\hc \hat{c}_{j\sigma},  \label{H-el}
\end{eqnarray}
where $\hbar$ and $m$ are the Planck constant and the electron mass, $\bvec{X} = (\bvec{X}_1, \ldots, \bvec{X}_N)$ and $\bvec{P}_i =
M_i\dot{\bvec{X}}_i$ are the coordinates and the momenta of the ions, respectively, $U(\bvec{X})$ is the ``bare'' ion-ion potential, and
$u(\bvec{x}; \bvec{X})$ is the effective (Kohn--Sham) potential. For simplicity, we assume one orbital per atom, so that the electron field
operator with spin projection $\sigma = \pm1/2$ is expanded as $\hat\psi_\sigma(\bvec{x}) \approx \sum_i\hat{c}_{i\sigma}
e_{i}(\bvec{x}-\bvec{X}_i)$ over an atomic-orbital basis set $\{ e_i(\bvec{x}) \}$, with the coefficients $\hat{c}_{i\sigma}$ being the electron
annihilation operators. The transfer integrals and on-site energies for these orbitals form the $t_{ij}$ matrix. The Ehrenfest equations for the
atomic coordinates give $\dot{\bvec{P}}_i \approx -\bvec\nabla_i U(\bvec{X}) + \bvec{F}^{(\text{el})}_i(\bvec{X})$, where the force acting on
atom $i$ due to electrons equals:
\begin{eqnarray}
    \bvec{F}^{(\text{el})}_i &=& -\sum_{\sigma} \int \bvec\nabla_i u(\bvec{x}; \bvec{X}) \cdot
                                                   \avg{\hat\psi_\sigma\hc(\bvec{x})\hat\psi_\sigma(\bvec{x})} \diff^3x
    \nonumber\\
    &\approx& \sum_{j,k,\sigma} \bvec{f}^{i}_{jk} \avg{\hat{c}_{j\sigma}\hc \hat{c}_{k\sigma}},
    \quad \nabla_i \equiv \pd / \pd \bvec{X}_i,  \label{F-general}
\end{eqnarray}
with the coefficients $\bvec{f}^{i}_{jk} = -\int e_j\cc(\bvec{x}) \bvec\nabla_i u(\bvec{x}; \bvec{X}) e_k(\bvec{x}) \diff^3x$. Note that the
total forces acting between the atoms contain the term coming from conduction electrons both in the presence and in the absence of the current
through the system. The coefficients $\bvec{f}^{i}_{jk}(\bvec{X})$, albeit not expressible in terms of $t_{ij}(\bvec{X})$, can be estimated
under reasonable assumptions. Namely, let us assume that (i)~the Kohn--Sham potential can be represented as a sum of \emph{even} contributions
$u_i(\bvec{x} - \bvec{X}_i)$ from individual atoms, (ii)~the basis orbitals, in turn, possess a definite parity, and (iii)~$u_i$ decays with a
characteristic spatial scale $1 / \zeta$ between atom $i$ and its neighbors. Then $\nabla_i u(\bvec{x}; \bvec{X}) \sim \zeta u_i(\bvec{x} -
\bvec{X}_i) \cdot \frac{\bvec{x} - \bvec{X}_i}{|\bvec{x} - \bvec{X}_i|}$, and $\bvec{f}^i_{ii}$ vanishes identically due to parity. The
nonnegligible coefficients that remain are $\bvec{f}^{i}_{jj} = -\nabla_i t_{jj}(\bvec{X}) \sim \bvec{f}^{i}_{ij} = \bvec{f}^{i\ast}_{ji} \sim
-\zeta \bvec{n}_{ij} t_{ij}$, where $\bvec{n}_{ij} \equiv \frac{\bvec{X}_j - \bvec{X}_i}{|\bvec{X}_j - \bvec{X}_i|}$, for the nearest neighbors
$j$ of site $i$ (denoted below as $j\in\text{NN}(i)$). The $\zeta$ parameter is naturally expected to be around the inverse atomic radius. The
resulting rough estimation of the force thus reads:
\begin{equation}\label{F-estimation}
    \bvec{F}^{(\text{el})}_i \sim -\zeta \!\! \sum_{j\in\text{NN}(i); \sigma} \!\! \bvec{n}_{ij} t_{ij} \avg{\hat{c}_{j\sigma}\hc \hat{c}_{j\sigma}
                                + \hat{c}_{i\sigma}\hc \hat{c}_{j\sigma} + \hat{c}_{j\sigma}\hc \hat{c}_{i\sigma}}.
\end{equation}

The expectation values in Eqs.~\eqref{F-general}, \eqref{F-estimation} are taken over a nonequilibrium many-body state describing a current of
conduction electrons. In order to describe it using NEGFs, we introduce a customary partition of the atomic sites into two ``leads'' $\LL, \,
\RR$ and the filament $\FF$ connecting them:
\begin{equation}
    \hat{H}^{\text{(el)}} = \sum_{S = \LL,\RR,\FF} \hat{c}_S\hc H_S \hat{c}_S + \sum_{S = \LL,\RR} \hat{c}_S\hc V_S \hat{c}_\FF,
\end{equation}
so that $\hat{c}_{\LL,\RR,\FF}$ denote columns with all annihilation operators in the leads or the filament, $H_{\LL,\RR,\FF}$ are the
Hamiltonian matrices of these three subsystems, and $V_{\LL,\RR}$ couple the filament to the two leads. The retarded NEGF of the filament is
introduced in a standard way,
\begin{equation}
    G_\FF(E) = (E - H_\FF - \Sigma_\LL(E) - \Sigma_\RR(E) + \ii\eta)^{-1}, \;
\end{equation}
where the self-energies $\Sigma_{\LL,\RR} = V_{\LL,\RR}\hc g_{\LL,\RR} V_{\LL,\RR}$ stem from isolated-lead Green's functions $g_{\LL,\RR}(E) =
(E - H_{\LL,\RR} + \ii\eta)^{-1}$, and $\eta \to +0$. The transmission probability in the form of Caroli is~\cite{Caroli}:
\begin{equation}\label{Caroli}
    \mathcal{T}(E) = \tr\{G_\FF\hc(E)\Gamma_\LL(E)G_\FF(E)\Gamma_\RR(E)\},
\end{equation}
with the broadening operators $\Gamma_{\LL,\RR} = \ii(\Sigma_{\LL,\RR} - \Sigma_{\LL,\RR}\hc) = 2\pi V_{\LL,\RR}\hc \rho_{\LL,\RR} V_{\LL,\RR}$,
where $\rho_{\LL,\RR}(E)$ are the spectral functions of the isolated leads.

Now, to find the expectation value in Eq.~\eqref{F-general}, we introduce ``incident'' wave functions $\varphi^{(n)}_{\LL,\RR}$ in the isolated
left or right leads, with energies $E_n$, so that the scattering state $\psi^{(n)}$ of the whole system is given by NEGFs: e.g., for the waves
coming from the left lead, $\psi^{(n)} \equiv \bigl(\psi_\FF^{(n)}, \; \psi_\LL^{(n)}, \; \psi_\RR^{(n)}\bigr) = \bigl(G_\FF V_\LL\hc
\varphi^{(n)}_\LL, \; (\idMatrix + g_\LL V_\LL G_\FF V_\LL\hc) \varphi^{(n)}_\LL, \; g_\RR V_\RR G_\FF V_\LL\hc \varphi^{(n)}_\LL \bigr)$. The
rest of the calculation is straightforward and takes its simplest form if both atoms belong to the filament, $j,k\in \FF$:
\begin{eqnarray}
    \avg{\hat{c}^\dagger_{j\sigma} \hat{c}_{k\sigma}} &=& \!\!\!\!\sum_{S = \LL,\RR; n} f(E_n - \mu_S) \psi^{(n)}_{k\sigma} \psi^{(n)\ast}_{j\sigma} \nonumber\\
    &=&\!\!\!\! \sum_{S = \LL,\RR; n} \int f(E - \mu_S)\diff{E}\; \bigl[ G_\FF(E) V_S\hc(E) \bigr.\nonumber\\
    &\cdot& \!\!\!\!\!\bigl.\{\varphi^{(n)}_S \!\!\otimes \!\varphi^{(n)\dagger}_S\delta(E - E_n)\} V_S(E) G_\FF\hc(E) \bigr]_{kj},
\end{eqnarray}
where $f(E-\mu)$ is the Fermi--Dirac distribution in the lead with chemical potential $\mu$. Observing that the term in braces gives nothing but
the spectral function $\rho_S(E)$, we readily yield:
\begin{equation}
    \avg{\hat{c}^\dagger_{j\sigma} \hat{c}_{k\sigma}} = \!\! \sum_{S = \LL, \RR} \int \frac{f(E - \mu_S)\diff{E}}{2\pi}\;
                                                                       \bigl[ G_\FF \Gamma_S G_\FF\hc(E) \bigr]_{kj}.
\end{equation}
Used together with expression~\eqref{F-general}, this yields the electron-induced, ``recoil'' force:
\begin{equation}\label{F-NEGF}
    \bvec{F}^{(\text{el})}_i(V)  =  \sum_{j,k} \bvec{f}^i_{jk} \!\!\!
                                 \sum_{S = \LL, \RR} \int_{E_0}^{\mu_S} \!\!\frac{\diff{E}}{\pi}\;\bigl[ G_\FF \Gamma_S G_\FF\hc(E) \bigr]_{kj},
\end{equation}
where $\mu_\LL = E_{\text{F}} - (1-\beta)e_0 V$ and $\mu_\RR = E_{\text{F}} + \beta e_0 V$ are the chemical potentials of the leads, $V$ is the
applied bias, $E_0$ and $E_{\text{F}}$ are the conduction band bottom and the Fermi level, respectively. The $\beta$ parameter controls a
possible asymmetric voltage drop in the leads attached to the constriction~\cite{Xue_review2019}; in what follows, we will assume the symmetric
case $\beta = 1/2$ unless otherwise specified. Note that the energy integral in Eq.~\eqref{F-NEGF} extends to the whole conduction band, unlike
the case of the Landauer--B\"uttiker formula for the conductance~\cite{Landauer,Buttiker}:
\begin{equation}\label{Landauer}
    G = \frac{G_0}{e_0 V} \int_{\mu_\LL}^{\mu_\RR} \mathcal{T}(E) \diff{E}.
\end{equation}
However, as we noted above, the zero-current interatomic interaction (e.g., the force field used in MD simulations) should include both the bare
term $-\bvec\nabla_i U$ and the one given by Eq.~\eqref{F-NEGF} \emph{integrated over $[E_0, \;E_{\text{F}}]$}. Therefore, the extra,
``renormalized'' force $\bvec{F}^{(\text{el})}_i(V) - \bvec{F}^{(\text{el})}_i(0)$ acting on the atoms at a nonzero bias originates from the
electronic states with $E \in [\mu_\LL, \;\mu_\RR]$, similarly to the Landauer formula.

It should also be noted that the mere momentum conservation law (which can be derived from the field-theoretic expression~\eqref{F-general})
gives the total force acting on the constriction aligned with a certain axis $z$:
\begin{eqnarray}
    F^{(\text{el})}_z \equiv \sum_i F^{(\text{el})}_{i,z} &=& \frac{2}{\pi} \sum_{S = \LL,\RR} \int_{E_0}^{\mu_S} \diff{E} \; \hat{z}_S
    \nonumber\\
                                                                &\times& \!\!\!\!\sum_{n: \;k_n^S(E) \in \Reals} (1-\mathcal{T}_n(E)) k_n^S(E),
    \label{F-total}
\end{eqnarray}
with the sum running through open channels in the leads with directions $\hat{z}_{\LL,\RR} = \mp1$, wave numbers $k_n^{\LL,\RR}(E)$, and
transmittances $\mathcal{T}_n(E)$. This force vanishes for a perfectly ballistic conductance ($\mathcal{T}_n = 1$); moreover, for identical
leads, the integral cancels out except the $[\mu_\LL, \;\mu_\RR]$ segment, analogously to the Landauer formula:
\begin{equation}\label{Ftotal-ala-Landauer}
    F^{(\text{el})}_z = \frac{2}{\pi} \int_{\mu_\LL}^{\mu_\RR} \diff{E} \!\!\!
    \sum_{n: \;k_n^\RR(E) \in \Reals} (1-\mathcal{T}_n(E)) k_n^\RR(E).
\end{equation}
For a single open channel at the Fermi surface and a small bias, $F^{(\text{el})}_z \approx \frac{2}{\pi} e_0 V k_{\text{F}} (1 - G / G_0)$,
where $k_{\text{F}}$ is the Fermi momentum, which suggests that forces measured in tenths of $\text{eV}/\Angstrom$ can act on the thinnest
filaments with their transmittances not close to unity.

Intuitively, it appears that similarly to the total force~\eqref{Ftotal-ala-Landauer}, the recoil forces~\eqref{F-NEGF} acting on individual
atoms should also be suppressed in the ballistic regime with the transmittance $\mathcal{T}(E)$ close to an integer, which, according to
Eq.~\eqref{Landauer}, is characterized with a quantized conductance. To test this physical intuition, below we consider an analytically solvable
case of a 1D chain of atoms, followed by an MD simulation of a nanofilament with realistic interatomic interactions and the quantum recoil on
top of them. While the former system is meant to serve as an idealized ``toy model'', letting us estimate the magnitude of the forces, their
dependence on the bias, and correlation with the transmittance, the latter simulation will give a definitive answer on how the recoil forces can
affect the evolution of a conducting filament and drive a resistive switching in a memristor.

\jetlsection{A quantum point contact in a 1D atomic chain.} Let us number the atoms of such a chain, schematically shown in Fig.~\ref{fig1}(b),
with $j = \pm1, \pm2, \ldots$ (right/left leads, respectively) and $j = 0$ (the single central atom comprising the ``filament''), so that at
zero bias, the equilibrium positions of the atoms are proportional to the lattice parameter $a$, $\bvec{X}_j = ja\bvec{e}_z$, along the $z$
axis. The nearest-neighbor transfer integrals involving the central atom are $t_{0,\pm1} = v_{\RR,\LL} \in \Reals$, while all the other
$t_{j,j\pm1} = t < 0$; let us also take the on-site energies $t_{jj} = \epsilon \delta_{j,0}$. As mentioned, within such a toy model, we are
interested in estimating the electron-induced force acting on the central atom, treating all the other, lead atoms as fixed, and the electron
transmittance of the chain. Now, working within the NEGF formalism described above, it is straightforward to find the ``incident-wave'' states,
$\varphi_{\LL,\RR;j}^{(k)} = \sqrt{2a/\pi} \sin jka$, $E_k = 2t\cos ka$, $0 < k < \pi/a$, and the self-energies $\Sigma_{\LL,\RR}(E) =
v_{\LL,\RR}^2 \int_0^{\pi/a} \diff{k}\; \phi_{\RR;1}^{(k)} \phi_{\RR;1}^{(k)\ast} / (E-E_k + \ii\eta) = (v_{\LL,\RR}^2/2t^2)(E - \ii\sqrt{4t^2 -
E^2})$. After that, the filament Green's function (a $1\times1$ matrix in our case) reads $G_\FF(E) = (E - \epsilon - \Sigma_\LL - \Sigma_\RR +
\ii\eta)^{-1}$, together with Eq.~\eqref{Caroli} giving a closed-form expression for the transmittance:
\begin{equation}
    \mathcal{T}(E) = \left[ \frac{((E-\epsilon)t^2 / v_\LL v_\RR - \chi E)^2}{4t^2-E^2} + \chi^2 \right]^{-1},
\end{equation}
where $\chi \equiv (v_\LL^2 + v_\RR^2) / 2 v_\LL v_\RR$. The conductance at small biases, i.e., near the Fermi surface $E \approx E_{\text{F}} =
0$, is $G = G_0 T(E_{\text{F}}) = G_0 / (\epsilon^2 t^2 / 4 v_\LL^2 v_\RR^2 + \chi^2)$. Careful evaluation of the total force acting on the
central atom ($j = 0$) gives an expression:
\begin{figure*}
    \begin{center}
        \includegraphics[width=0.95\textwidth]{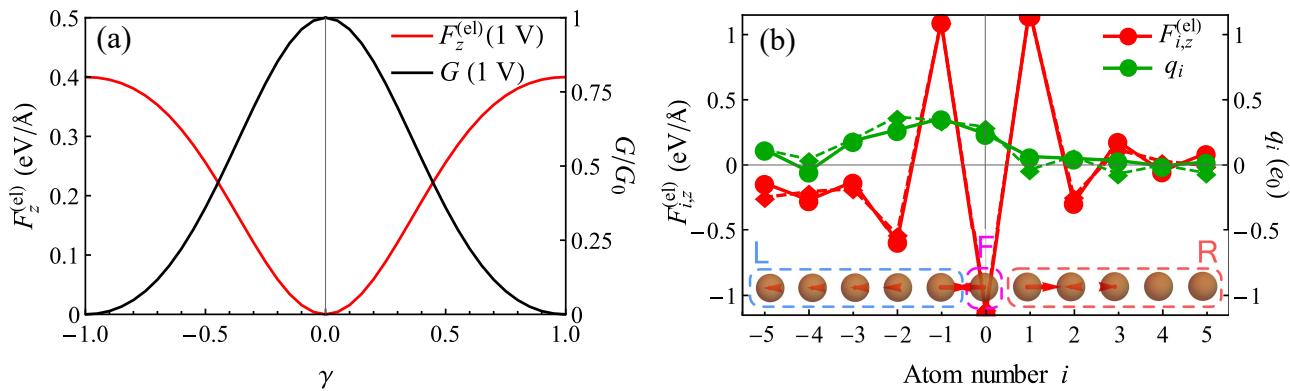}
    \end{center}
    \caption{(a) Conductance and total recoil force acting on a 1D atomic chain with a point contact with $v_{\LL,\RR} = (1\pm\gamma)t$
    at a bias $V = 1~\text{V}$; (b) Forces acting on individual atoms and their partial charges in the same chain at $V = 0 \text{ and } 1~\text{V}$
    (solid and dashed lines, respectively), for $\gamma = 0.4$ corresponding to transmittance $\mathcal{T}(E_{\text{F}}) \approx 0.5$.
    In both panels, $t = -1.8~\text{eV}, \; \epsilon = 0$, the interatomic distance $a = 2.5~\Angstrom$,
    and $\zeta = 1/a$. Note also that at zero bias, the total force in panel~(a) vanishes.}
    \label{fig1}
\end{figure*}
\begin{gather}
    F^{\text{(el)}}_{0,z}(V) \approx -2\zeta \int \sum_{p=-1}^5 C_p \Phi^{(p)}(E)\diff{E}, \\
    \Phi^{(p)} \equiv (v_\RR^p f(E - \mu_\RR) - v_\LL^p f(E - \mu_\LL)) \times \frac{-\IIm g_\RR^{1,1}(E)}{\pi}, \nonumber\\
    C_{-1} = -v_\LL^3 v_\RR^3 C_5, \quad C_0 = -v_\LL^2 v_\RR^2 E C_5, \quad C_4 = E C_5,          \nonumber\\
    C_1 = 1, \quad C_2 = 2 \RRe G_\FF, \quad C_3 = 2 \RRe(g_\RR^{1,1} G_\FF),                                       \nonumber\\
    C_5 = |G_\FF|^2 / t^2, \nonumber
\end{gather}
where $g_\RR^{1,1}(E) = (E - \ii \sqrt{4t^2 - E^2}) / 2t^2$ and the approximation~\eqref{F-estimation} is used. One immediately observes that at
zero bias, the force contains a prefactor of $v_\RR - v_\LL$ coming from $\Phi^{(p)}$ and thus vanishes for a symmetric chain, $v_{\LL,\RR} =
t$. Below we quote the leading order of the zero-bias force in $v_\RR - v_\LL \to 0$ (the on-site energy $\epsilon$ does not contribute within
the linear order):
\begin{eqnarray}\label{F-approx-zero}
    F^{\text{(el)}}_{0,z}(0) &\approx& \frac{2\zeta (v_\LL - v_\RR)}{\pi t^2}
                                                 \int_{E_0}^{E_{\text{F}}} \frac{E^2 + 2E t - t^2}{\sqrt{4t^2 - E^2}} \diff{E} \nonumber\\
    &=& (1 + 8/\pi) \zeta (v_\LL - v_\RR),
\end{eqnarray}
where we have made use of the fact that for our chain at half-filling, $E_0 = 2t < 0$ and $E_{\text{F}} = 0$. At a finite but small bias,
assuming that $v_{\LL,\RR} = (1 \pm \gamma) t$, $\gamma \to 0$ (which could be a result, e.g., of a deviation of the central atom from the
origin along the $z$ axis), one can find the ``renormalized'' force:
\begin{equation}\label{F-approx-finite}
    F^{\text{(el)}}_{0,z}(V) - F^{\text{(el)}}_{0,z}(0) \approx \frac{\zeta e_0 V}{\pi t}
                                                                \Bigl[ \Bigl(\beta-\frac12\Bigr)(v_\LL - v_\RR) - \epsilon\Bigr].
\end{equation}
In fact, it is the reflection symmetry mapping the single-atom ``filament'' onto itself and resulting in $F^{\text{(el)}}_{0,z} \to
-F^{\text{(el)}}_{0,z}$, $V \to -V$, $\beta \to 1-\beta$, $v_{\LL,\RR} \to v_{\RR,\LL}$, which forbids the contribution proportional to $V(v_\LL
- v_\RR)$ for a symmetric voltage drop ($\beta = 1/2$). In the next section, we will see that for larger filaments with their different ends
coupled to different leads, this cancelation does not occur, leading to pronounced finite-bias recoil effects even for symmetric geometries. An
estimation with the interatomic distance $a \sim 2.5~\Angstrom$, $\zeta \sim 1/a$, $V \sim 1~\text{V}$, $t \sim -2~\text{eV}$, and $v_\LL -
v_\RR, \; \epsilon \sim 0.5~\text{eV}$ gives the zero-bias force of about $0.7~\text{eV}/\Angstrom$ and the finite-bias contributions around
$0.02\text{--}0.03~\text{eV}/\Angstrom$. While the latter figure is quite small, the former one is considerably large at the scale of
interatomic interactions and agrees with our above expectations of the total recoil force~\eqref{Ftotal-ala-Landauer}.

It is also interesting to note the effect of the applied bias on the stability of the position $Z_0$ of the central atom. Indeed, assuming that
the transfer integrals involving this atom are $v_{\LL,\RR} \sim t e^{\mp\zeta Z_0}$ and $\epsilon = 0$, one has the equilibrium position $Z_0 =
0$ with both the electronic force~$F^{\text{(el)}}_{0,z}(V)$ and the ionic contribution~$-\pd U / \pd{Z_0}$ vanishing (due to $v_\LL = v_\RR$
and the symmetry, respectively). If the central atom infinitesimally deviates from $Z_0 = 0$, the total force acting on it becomes $F_{0,z}(V;
Z_0) \approx -2\zeta^2 t Z_0 \bigl[ 1 + 8 / \pi + (\beta - 1/2) e_0 V / \pi t \bigr] - (\pd^2 U / \pd Z_0^2) Z_0$. Now, taking the ion-ion term
in the form of a screened Coulomb interaction with the nearest neighbors, $U = q_{\text{eff}}^2\bigl( |Z_0+a|^{-1} + |Z_0-a|^{-1} \bigr) +
(\text{terms independent of }Z_0)$, where $q_{\text{eff}}$ is the screened ion charge, we arrive at the following hessian:
\begin{equation}\label{hessian}
    \frac{\pd F_{0,z}(V; Z_0)}{\pd Z_0} = -\frac{4 q_{\text{eff}}^2}{a^3}
                                                - 2\zeta^2 t \Bigl[ 1 + \frac{8}{\pi} + \frac{(\beta - 1/2) e_0 V}{\pi t} \Bigr].
\end{equation}
While we assume that $\pd F_{0,z}/\pd Z_0 < 0$ at zero bias to ensure the stability of the chain before applying the voltage, the latter can
make it unstable if $\sgn{V} = \sgn(1/2-\beta)$, above the critical value:
\begin{equation}
    |V| > V_{\text{crit}} = \frac{\pi |t|}{e_0 |1/2-\beta|} \left[ \frac{2 q_{\text{eff}}^2}{\zeta^2 a^3 |t|} - 1 - 8/\pi \right].
\end{equation}
In particular, estimations show that the destabilizing zero-bias term in brackets $1 + 8/\pi \approx 3.5$ can well compete with the stabilizing
Coulomb one $2 q_{\text{eff}}^2 / \zeta^2 a^3 |t| \approx 5.7(q_{\text{eff}}/e_0)^2$ for partially screened charges. For example, for
$q_{\text{eff}} = 0.8 e_0$ and $\beta = 0$, application of a voltage above $V_{\text{crit}} \approx 1.7~\text{V}$ results in a localized
counterpart of the Peierls instability affecting the central atom's position~\cite{Peierls}. In fact, this voltage-induced instability could
become even more pronounced for more complex filament geometries, in particular, not exhibiting a reflection symmetry.

To complement the discussed analytical estimations, Fig.~\ref{fig1}(a) demonstrates the total recoil force in a chain with $\epsilon = 0$ versus
the conductance (i.e., transmittance), while Fig.~\ref{fig1}(b) resolves the force and the charges at different atoms around the quantum point
contact. Note that according to Eq.~\eqref{F-general}, the recoil forces arise from interactions of the ions with Friedel oscillations of the
electron density~\cite{Friedel}, thus, these forces should decay into the bulk together with the density oscillations. The amplitude of the
forces, as we observe in Fig.~\ref{fig1}(b), is large enough to perturb the geometry of the contact and either push it toward a configuration
with a transmittance close to unity (when the recoil is negligible), or to facilitate ion migration. In this regard, it is worth noting that
even for a constriction $\sim 10~\Angstrom$ (four atoms) long, the conventional, electrostatic force driving ion migration in memrsitors should
roughly be $\sim 0.1~\text{eV} / \Angstrom$ at the voltage of $1~\text{V}$---compared with $\sim 1~\text{eV} / \Angstrom$ in
Fig.~\ref{fig1}(b)--- which highlights the potentially important role of the recoil forces in the evolution of thinnest conductive filaments in
memristive devices. Note in this connection that, although we have mentioned above that the recoil force should be renormalized in some way to
avoid double counting upon its addition to the ``bare'', classical ion-ion interaction $U(\bvec{X})$, the oscillations in Fig.~\ref{fig1}(b)
reflect the quantum nature of the force and cannot be renormalized out after subtraction of the interference-free, classical term. Let us now
switch to a more realistic numerical simulation of the effect of the recoil forces, to complement the outcomes of the present paragraph (in
particular, to observe the \emph{renormalized} recoil forces exceeding $0.1~\text{eV}/\Angstrom$).

\begin{figure}
    \begin{center}
        \includegraphics[width=0.9\columnwidth]{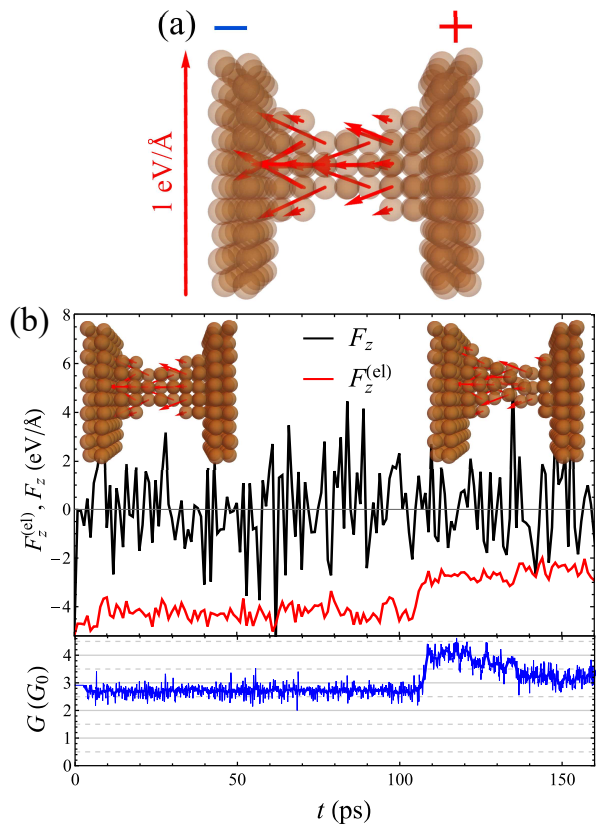}
    \end{center}
    \caption{Effect of the electron recoil forces on the statics and dynamics of the Cu filament:
    (a)~a pre-generated symmetric filament with a pure fcc geometry and the forces acting on its atoms at $V = 1~\text{V}$;
    (b)~recoil and total (recoil+EAM) forces acting on the filament (above), and its conductance (below) during the filament evolution
        at $T = 300~\text{K}$.
    The arrow on the left in panel (a) shows the scale corresponding to a force $1~\text{eV}/\Angstrom$;
    the voltage drop is symmetric ($\beta = 1/2$).}
    \label{fig2}
\end{figure}

\jetlsection{Simulation of a Cu filament with realistic interatomic forces.} Of course, the filaments actually forming in memristors are neither
one atom wide, as in Fig.~\ref{fig1}(a), nor centrosymmetric: they result from stochastic ion migration, e.g., during the so-called
electroforming of the memristor, which precedes its normal operation~\cite{Lee_review}. Therefore, the effect of the recoil forces on such,
general filaments at room temperature may be beyond the conclusions drawn from the above 1D toy model at $T = 0$. To simulate the dynamics and
study the recoil-driven ion migration along such filaments, we developed a custom MD code implementing the embedded-atom model (EAM) of the
interatomic Cu-Cu forces \cite{Sheng_EAM} and used a pre-generated typical initial geometry of a conductive filament between two ideal fcc
electrodes [Fig.~\ref{fig2}(a)], further warmed up to room temperature to obtain a more representative configuration. Following the adiabatic
approximation, the recoil forces were calculated for instantaneous ion positions using Eq.~\eqref{F-NEGF} and then renormalized by subtracting
their zero-bias values. Approximation~\eqref{F-estimation} was used for the force coefficients $\bvec{f}^i_{jk}$ with $\zeta = 1/a$, where $a =
2.54~\Angstrom$ is the nearest-neighbor distance in the bulk fcc copper. When parameterizing the tight-binding model~\eqref{H-el}, the
differences between the on-site energies of the atoms were neglected, while the transfer integrals were evaluated as $t_{ij} =
-1.8~\text{eV}\times e^{-\zeta |\bvec{X}_i - \bvec{X}_j|}$, approximating the corresponding DFT results for $4s$ states in Cu dimers. As for the
Green's functions $g_{\LL,\RR}$ of the semi-infinite leads, explicit Fourier-series expressions in the nearest-neighbor approximation were used
for them, allowing one to avoid known poor conditioning issues~\cite{PoorConditioning}. Following a typical MD simulation methodology, we first
minimized the energy of the system at absolute zero, then warmed it up to $T = 300~\text{K}$, and finally ran a $150~\text{ps}$ production MD,
using the Langevin thermostat. Being quite time-consuming, evaluation of the recoil forces was done each $0.1~\text{ps}$ of the simulation,
while the Langevin equation was integrated with a $1~\text{fs}$ timestep.

The recoil forces calculated for the initial geometry [Fig.~\ref{fig2}(a)] exhibit a strong asymmetry, despite the atomic positions for this
configuration were intentionally chosen at the sites of an ideal fcc crystal. The asymmetry is caused by the bias $V = +1~\text{V}$ applied to
the anode (the right electrode in the figure): in agreement with Eq.~\eqref{F-general}, the forces point toward the cathode, which has a higher
electron density due to the excess electron waves arriving there from the lead. The forces are measured in several tenths of
$\text{eV}/\Angstrom$, as in the case of the 1D atomic chain. As finite temperature is switched on, the conductance, initially being very close
to $3G_0$, develops fluctuations around its quasisteady state and further features a sharp resistive switching around $t = 110~\text{ps}$
[Fig.~\ref{fig2}(b)]. During all this time, the total recoil force remains directed toward the cathode, facilitating ion migration in this
direction; in agreement with our above analysis, the force drops down when new conduction channels open. The result of the ion migration is seen
in the initial and the final states of the filament presented in Fig.~\ref{fig2}(b): a whole atomic layer near the anode gets rarified, with the
atoms adding to the central part of the filament and thus enhancing the overall conductance. Note that, though the migration has the same
direction as that of $\text{Cu}^+$ ions driven by the electric field, the latter is absent in our simulations, and the ion transport occurs
solely due to momentum transfer from conduction electrons we (perhaps somewhat misleadingly) refer to as the ``recoil'' here. Thus, we conclude
that a specific type of ion electromigration occurring due to a nonequilibrium quantum current of charge carriers can drive resistive switching
in the atomic-scale-filaments regime.

\jetlsection{Discussion.} Finally, a couple of words should be said on the applicability of the approximations used above. First, the
tight-binding model is a useful, easily parameterizable framework, but is more suitable for describing semiconductors, rather than metals. At
the same time, in atomic-thickness filaments, the screening effect may be weaker than that in the bulk metal, with the real physics lying
between the tight-binding model and the free-electron one considered recently in Ref.~\cite{Kharlanov2022}. Second, in the context of these two
models, a nontrivial thing to be noted regards the contributions of individual atoms to the total Kohn--Sham potential $u(\bvec{x}; \bvec{X})$:
the additive representation of the latter used in the present work clearly deserves further study and generalization. Third, in the above, we
treated electron-current-induced forces in the mean-field fashion: this approximation rests on the fact that at the voltage of $1~\text{V}$,
around $5 \times 10^{14}$ electrons per second pass through a filament with conductance $G_0$, so that the time it takes a copper atom to shift
by $0.1~\Angstrom$ is enough for around 15, i.e., quite a lot of electrons to pass through the filament. Finally, for transition metals,
including copper, localized $d$~orbitals are important, so that to achieve a better numerical accuracy, a multiscale \emph{ab~initio}+MD
methodology appears more appropriate. For these further steps, however, the calculations within our simplified model may serve as a
proof-of-principle study.

Last but not least, we would like to outline the novelty of the presented study. The main result is the quantum recoil-driven resistive
switching mechanism demonstrated above, which works independently of the conventional ion migration in the electrostatic field. Namely, the
forces acting on the ions and mediated by electrons get imbalanced due to the difference in the chemical potentials of the two leads, providing
an additional driving force for ion migration. Note that unlike the continuous-medium approach applied in Ref.~\cite{Kharlanov2022} to
atomic-thickness filaments, the use of the more realistic atomistic potentials does not lead to the \emph{transversal} capillary-bridge
instability of the filament due to surface tension forces~\cite{CapillaryInstabilities}; instead, the ion migration proceeds \emph{along} the
filament. Clearly, further simulations with larger and/or asymmetric filaments could reveal more interesting results, however, we leave them
beyond the present work because of their higher computational demand. Also, though the very tools we have used for describing the quantum recoil
forces---the NEGF formalism and the tight-binding model---are quite conventional and well-established \cite{Dundas_NNano2009, Todorov_PRB2010,
Do_NEGFreview, Caroli}, the forces are, to the best of our knowledge, included in a dynamical description of the conducting filament for the
first time.

\jetlsection{Conclusion.} To summarize, we have studied the effect of the ``recoil'' forces between the conduction electrons and the ions
comprising a conductive filament on the geometry of the latter. As both analytical estimations and the MD simulation have revealed, the effect
of the recoil should be quite pronounced for conductive filaments several atoms wide, like those formed in memristors exhibiting the conductance
quantization effect. Unlike the classical volume and surface forces acting on the filament, the quantum recoil is intrinsically related to the
transmittance and additionally destabilizes the filaments with the non-integer conductance. Our study thus provides a framework and highlights
the importance of taking into account the charge-carrier recoil in the rational design of current and next generations of memristive
nanodevices.

O.K. thanks Anton Minnekhanov for fruitful discussions. The work was supported by the Russian Foundation for Basic Research (Project
No.~20-07-00696). This work has been carried out using computing resources of the federal collective usage center Complex for Simulation and
Data Processing for Megascience Facilities at NRC ``Kurchatov Institute''.

\end{document}